\documentclass[letterpaper,twocolumn,10pt]{article}
\usepackage{usenix-2020-09}

\usepackage{todonotes}
\usepackage{graphics} 
\usepackage{epsfig} 
\usepackage[fleqn]{amsmath} 
\usepackage[nolist]{acronym}
\usepackage{multirow}
\usepackage[ruled,vlined,commentsnumbered]{algorithm2e}
\usepackage{color}
\usepackage{listings}
\usepackage{subfigure}

\usepackage{algorithm2e}

\usepackage{amssymb}
\usepackage{amsthm}

\usepackage{pgf}
\usepackage{tikz}
\usetikzlibrary{arrows,automata}

\newtheorem*{researchquestion}{\textbf{Research Question}}
\newtheorem{hyp}{Hypothesis}
\newtheorem{nullhyp}{Null Hypothesis H$_0$}

\begin{document}
\title{\textbf{ Detecting Ransomware Execution in a Timely Manner }}


\author{\textbf{Anthony Melaragno}
    \\
	Cyber Sciences\\
		United States Naval Academy\\
		Annapolis, Maryland \\
		melaragn@usna.edu
	\and
	\textbf{William Casey}
	\\
	Cyber Sciences\\
	United States Naval Academy\\
	Annapolis, Maryland \\
    wcasey@usna.edu}

\maketitle

\begin{abstract}
Ransomware has been an ongoing issue since the early 1990s.  In recent times ransomware has spread from traditional computational resources to cyber-physical systems and industrial controls.  We devised a series of experiments in which virtual instances are infected with ransomware. We instrumented the instances and collected resource utilization data across a variety of metrics (CPU, Memory, Disk Utility).  We design a change point detection and learning method for identifying ransomware execution. Finally we evaluate and demonstrate its ability to detect ransomware efficiently in a timely manner when trained on a minimal set of samples.  Our results represent a step forward for defense, and we conclude with further remarks for the path forward.
\end{abstract}

\begin{lstlisting}[basicstyle=\bfseries]
Machine Learning Ransomware Identification 
Change Detection
\end{lstlisting}


\begin{acronym}
\acro{ACL}{Access Control List}
\acro{CECSR}{Computer Engineering and Cyber Security Research}
\acro{AI}{Artifical Inteligence}
\acro{DevOps}{Development Operations}
\acro{DoS}{Denial of Servcie}
\acro{DDoS}{Distributed Denial of Service}
\acro{GUI}{Graphical User Interface}
\acro{IO}{Input-Output}
\acro{ML}{Machine Learning}
\acro{ONR}{Office of Naval Research}
\acro{RAID}{Redundant Array of Inexpensive Disks}
\acro{USB}{Universal Serial Bus}
\acro{USNA}{United States Naval Academy}
\end{acronym}

\section*{Introduction}
The term {\it ransomware} refers to software which is designed to ransom access to user data, resources and information by means of manipulating access control to deny standard access modes.  
In the most common attack, ransomware takes user data hostage (most often with encryption), then offers a ransom price for the encryption key so the user can regain data access.

Ransomware is a considerable problem to contemporary cyber security \cite{cohen_2021}, studies have estimated the damages due to ransomware on course to exceed \$265B this year alone.   
According to \cite{ransometax} there has been a continuous and substantial rise of ransomware resulting in large monetary loss. The origin of ransomware has an interesting progression staring in 1989 with a software program designed to manage personal medical information and risk factors associated to HIV/AIDS written by Dr. Joseph Popp~\cite{popp}.  The program would execute normally until the $90^{th}$ execution, presumably reached only when the program proved useful within medical clinics, after which the behavior switched modes by encrypting file names for data (but not contents), until an upgrade fee was paid.  In this way the basic attack form of ransomware was completed by holding clinic data at risk, withholding the data until a ransom was paid.  Security researchers of  {\it cryptovirology} \cite{young1996cryptovirology} in the mid to late 1990's realized that a stronger attack is formed by encrypting not only the filenames but also file contents and  identified the current form of ransomware attack.  In 2012 an early ransomware attack called MoneyPak preyed upon unscrupulous users of pirated content, by encrypting user data while falsely claiming FBI authorities -- likely used to add credibility to the ransomware recovery option which was at that time not widely understood.  
In 2013 the Cryptolocker ransomware put ransomware on the map for most computer security communities, owing mainly to its broad and serious threatening nature to cyber security.  This turning point maybe due mainly to Cryptolocker's wide distribution facilitated by the Zues, Zbot, Gameover Botnet.  Despite weaknesses in the hand rolled encryption; the Cryptolocker ransomware is believed to have extorted over 3 million from victims before its 2013 take-down --  a first of its kind international police operation of multiple law enforcement entities including FBI and Interpol.  
Since Cryptolocker, ransomware has exploded in growth, with thousands of variants, including recent variants that include aspects of data destruction (Mongolock) and time pressure (Jigsaw) which erases a batch of files every minute and a thousand files for each reboot.
Additionally ransomware attacks have mixed with criminal hacking activities, and have reemerged as cyber attacks which are able to hold a critical service at risk, such as was done to Baltimore's Computer aided dispatch system (911 emergency dispatch system), and Bristol Airport's Flight Information Display System.   
Evan while ransomware activities usually fall within the criminal code of every country (i.e., blackmail or extortion), the current state of international law is forestalled with no effective ability to extradite indited or known actors.
In particular non-signatory countries of the Budapest Accord (the first diplomacy related to cybercrime) offer safe harbor to the ransomware business.  As of 2017 ransomware has become a worldwide issue hitting nearly 100 countries world wide~\cite{massiveransomware}.  

Generally speaking the types of ransomware (discussed in ~\cite{ransometax}) can range from password protected zip files to using cryptography in attacks.  There are mitigation and best practice techniques that can be used to aide in prevention and data recovery:
\begin{itemize}
	\item[]\textbf{Data Redundancy} performing data redundancy techniques such as backing up data to the cloud, locally to a \ac{RAID} or other physical storage media such as a \ac{USB} drive.
	
	\item[]\textbf{Security Applications} such as antivirus applications  can provide protections from ransomware running.  These tools can range from using signature files to using \ac{ML} and \ac{AI} techniques to identify ransomware.  
	
	\item[]\textbf{Software Maintenance} performing routine data maintenance such as updating and patching of software.
	
	\item[]\textbf{Screening Email / Phishing} screening incoming email for phishing attempts to filter suspect emails from vulnerable users.
	
	\item[]\textbf{Access Control \& Permissions} enforcing access control and permissions has the potential to prevent ransomware and malware generally from potentially accessing files and directories that they do not need access.  
	
	\item[]\textbf{Application Whitelisting} creating a whitelist of screened acceptable application which can only install and need specific user permissions to execute.
\end{itemize} 

Each of the listed mitigation techniques are valuable and should be employed to prevent data from being held hostage.  Each of the mitigation techniques have drawbacks.  Data Redundancy if infrequent could contain stale backup data. Security Applications, can not prevent zero day type of attacks due to dated signature files. Phishing emails if properly crafted could evade detection.  Access Control \& Permissions, Application allow lists would have to take into account both users and applications \ac{ACL}.  Maintenance of an \ac{ACL} can be problematic for large networks and systems.

\section*{Current Approach and Deficiencies}
\label{sec:currentapproach}

Currently the state-of-the-art malware detection~\cite{kharaz2016unveil,kharraz2017redemption, mehnaz2018rwguard} provides an initial starting point for ransomware detection and prevention. The systems described in~\cite{kharaz2016unveil,kharraz2017redemption, mehnaz2018rwguard} are host-based detection and prevention systems termed RAID, not to be confused with 
the standard notion of RAID systems
as defined in 1987 by David Patterson, Randy Katz and Garth A. Gibson. 

\begin{figure}[!h]
	\begin{center}
		\includegraphics[width=0.8\linewidth]{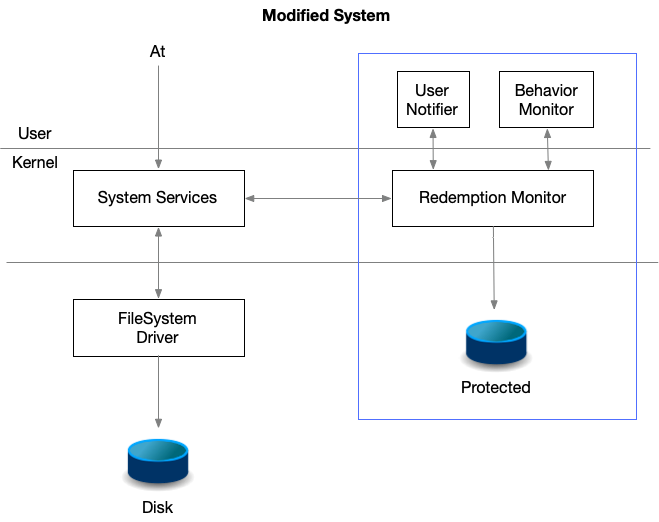}
	\end{center}
	\caption{"RAID" Detection and Deterrence Architecture~\cite{kharaz2016unveil,kharraz2017redemption}}
	\label{fig:raidridic}
\end{figure}

The system defined in~\cite{kharaz2016unveil, kharraz2017redemption} and redrawn in figure~\ref{fig:raidridic} contains the ability to intercept file IO operations via a kernel module which performs inspection and characterization of behavior of file~\ac{IO} frequency and entropy.  
The system also performs a backup of the corresponding file in a protected area as shown in figure~\ref{fig:raidridic}.  Essentially, the main idea is behaviour monitoring with a layer of redundancy and recovery to provide backup and restoration for user data.    

We summarize the main and variations of ideas of exiting approaches as: 

\begin{itemize}
	\item[]\textbf{Threshold Based Detection}: sets a statically based IO counter  to perform read write frequency analysis.  The detector logs the IO and performs an analysis of how quickly items are read to and from the behavior monitor~\cite[\S4.1]{kharraz2017redemption}.  The kernel module keeps track of the write score and makes a hard determination if it is ransomware. 
	
	\item[]\textbf{Database Like Subsystem}: referencing~\ref{fig:raidridic} the system adds and commits files to the protected area based on a notification from the redemption monitor~\cite[Data Consistency\S4.3]{kharraz2017redemption}. Once committed to the database the system uses that as a redundant backup to restore potentially ransomed files. 
	
	\item[]\textbf{Entropy Based Detection}: entropy is used as a factor for frequent file IO with a divergence~\cite{61115} measure similarly to ~\cite[\S5.1]{kharraz2017redemption} and the use of entropy to detect \ac{DoS} based attacks.  
	
	\item[] \textbf{File IO Based Systems}: detection and prevention systems such as ShieldFS~\cite{10.1145/2991079.2991110} uses a methodology to self adapt based on learned behavior of a trained system using file based IO requests and compares it to a file IO model.  The model in ShieldFS~\cite{10.1145/2991079.2991110} is specific to a windows based architecture.            
\end{itemize}

\subsection*{Current Approach Deficiencies}

The approach described 
in~\cite{kharaz2016unveil, kharraz2017redemption} 
and designed specifically for Microsoft Windows based kernel modules may be encumbered by portability issues when taken to other operating systems such as Apple OS-X and Linux operating Systems where ransomware remains a threat.   
Still its unclear if and how the ideas are general enough to be ported from one operating system to another.   

To address these problems we focus on higher level behavioral factors as the subjects of ensemble learning methods.  
By proceeding in this way, more complex models can be developed from smaller ones that pinpoint single behaviors.  Additionally we believe that a more general and promising route to higher assurance detection can be made more portable and effective.  
However efficient a detection paradigm can be challenges will remain due to the inherent adversarial chase and attacks are likely to evolve to detection.  
We therefore consider the
%
following list of issues as the setting for future detection and adversary evasion, we illustrate these using current art:

\begin{enumerate}
	\item[] \textbf{By-passible Threshold}: The threshold (i.e. malice score), once learned by the ransomware attacker, the ransomware can be designed to gauge the threshold to perform read and writes that are not indicative of malware.  In addition, since it is a kernel module the ransomware itself prior to execution could look for the kernel module extension and attack and disable the system
	\item[] \textbf{Access Frequency}: There could be scenarios in which the access frequency will lead to greater amounts of false positives which would render the system useless since a user would most likely turn off the module.  The adversary can tune their attack to thwart IO detection
	\item[] \textbf{Resource Impact}: Having essentially a secondary file system will inherently cause resource and synchronization issues. These issues can be an avenue of a secondary attack vector.  If the attacker is able to gain kernel access they could undermine the entire underpinning of the system.  Additionally, the structure of the protected area seems to be repetitive of an actual traditional \ac{RAID} system.  An actual \ac{RAID} would provide greater backup, restore (storage) capability then that which has been shown in figure~\ref{fig:raidridic}. 
	\item[] \textbf{Entropy Analysis}: There has been research in deceiving entropy based detection systems~\cite{10.1117/12.2054434} for \ac{DoS} based attacks by studying access based on calculating background entropy and use that to thwarting the detector.  A similar methodology can be used against the proposed architecture in figure~\ref{fig:raidridic}~\cite{kharraz2017redemption, kharaz2016unveil}.  The attacker can monitor background IO requests then create then estimate a distribution function and ensure that the encryption requests would fit the distribution function as they are attacking the proposed system~\cite{kharraz2017redemption, kharaz2016unveil}.  Essentially, according to~\cite{10.1117/12.2054434} it would neuter a entropy based detection system.  
\end{enumerate}

In this paper we will design and investigate the performance characteristics of a \ac{ML} method to aid in the identification of ransomware. Our contributions correct for possible deficiencies in ~\cite{kharraz2017redemption, kharaz2016unveil}. 
\begin{enumerate}
	\item The design of a light weight system monitor for general features. 
	\item A model for change detection caused by ransomware execution which generalizes across execution conditions.  
	\item An ML based construction method for decision procedure (we term decision gate) that requires only a single-shot or very few examples in training to render an Adaboost decision stumps classifier for relevant change points in monitored data, the resulting classifier is shown to accurately fire within a few seconds of ransomware execution. 
\end{enumerate}
The rapidness of the online detector may further enable the feasibility of dynamic access controls and permission enforcement that could potentially seek to contain a detected ransomware attack. 

\begin{table*}[!ht]
	\caption{Experiment Set Decomposition}
	\label{tbl:experimentsetdecomp}
	\centering
	\begin{tabular}{l||l}
		\textbf{Set Expression} & \textbf{Experiment Description} \\ \hline
		\begin{math}
		H_{0} : [1, 40] \rightarrow \{ t_i, t_s, t_b \} : t_e \rightarrow 
		\begin{cases} 
		t_i \text{ idle }  \\
		t_s \text{ installation $\parallel$ compression  } \\
		t_b \text{ system shutdown}
		\end{cases}
		\end{math} & Null Hypothesis : Parallel Execution Excluding Ransomware \\ \hline
		
		\begin{math}
		H_{1a} : [1, 40] \rightarrow \{ t_i, t_s, t_b \} : t_e \rightarrow 
		\begin{cases} 
		t_i \text{ idle } \\ 
		t_s \text{ install $\parallel$ compress $\parallel$ ransom } \\
		t_b \text{ system shutdown}
		\end{cases}
		\end{math}	 
		& Hypothesis Testing : Parallel Ransomware Execution \\ \hline
		
		
		\begin{math}
		H_{1b} : [1, 40] \rightarrow \{ t_i, t_s, t_b \} : t_e \rightarrow 
		\begin{cases} 
		t_i \text{ idle } \\ 
		t_a \text{ installation, compression} \\
		t_r \text{ ransomware} \\
		t_b \text{ system shutdown}
		\end{cases}
		\end{math}	 
		& Hypothesis Testing : Sequential Execution \\ \hline
	\end{tabular}
\end{table*}

\section*{Research Question and Hypothesis}

\begin{researchquestion}
	Can Change Detection be Used to Efficiently Detect Ransomware Execution in a Timely Manner?
\end{researchquestion}

We decompose the research question into the following hypotheses: accuracy, performance, and features.  Experiments are constructed, automated specifically to test and reveal aspects of resources of a system under attack.
 
\begin{hyp}
\label{hyp:accuracy}
$H_1$ Machine Learning learning training upon changes in system resources (i.e. memory, cpu, and disk utilization) utilizing change point detection will be able to detect the execution of ransomware with sufficient accuracy to allow for preventative actions. 
\end{hyp}

\begin{nullhyp}
\label{null:hypaccuracy}
$H_0$ A machine learning algorithm will not have sufficient accuracy to detect ransomware execution even with the application of change point detection to detect changes in system resources.
\end{nullhyp}

We decompose Hypothesis $H_1$ into two separate hypothesis to add in randomization of  application execution and ransomware execution in a sequential as well as ransomware executing in parallel with applications.  Table~\ref{tbl:experimentsetdecomp} summarizes the experimental set decomposition.

\section*{Experiment}
A series of experiments were conducted where 
ransomware~\cite{gonnacry} was installed into a virtual environment using both headless and \ac{GUI} enabled LINUX operating systems.  \ac{DevOps} was used to automate the experimentation in a series of tests. The experiments are set for normal execution, which the user is using the system without any execution of ransomware and represents the control.  Timed execution, which the virtualized environment is being run with and without ransomware.  The virtual machine starts and operates without ransomware for 10 minutes then executes with ransomware for 20 minutes. Immediate execution where ransomware is started immediately once the virtual environment has been started and random execution. Shown in  table~\ref{tbl:experimentsetdecomp} Experiment Set Decomposition there are 4 types of experiments that were run to prove or disprove the hypothesis using hypothesis testing we will reference the table when describing the test setup.

\subsection*{Experimental Setup and Inital Observations}

The experiments  set expressions described in table~\ref{tbl:experimentsetdecomp} drove the experimentation.  Forty experiments were run with data being collected every 0.5 seconds for a total of 30 minutes for each expression on 4 instances per experiment.  Within the 30 minutes execution times were randomly chosen for the events.  A single experiment represents a set of 4 operating systems executing with set random times within a 30 minute time window prior to destruction.  The experiment recalculates random times and executes again until all 40 experiments are concluded thereby having statistical significance.  These experiments are:
\begin{itemize}
	\item [] H$_0$ Null Experiment | Parallel Execution Excluding Ransomware
	\item [] $H_{1a}$ Parallel Ransomware Execution
	\item [] $H_{1b}$ Idle with Sequential Ransomware Execution
\end{itemize}

\begin{figure}[!h]
	\begin{center}
		\includegraphics[width=1\linewidth]{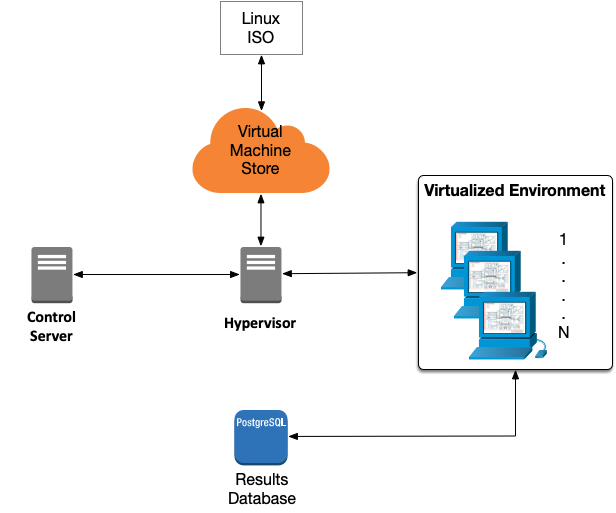}
	\end{center}
	\caption{Experimental Setup}
	\label{fig:testsetup}
\end{figure}

The experimental setup as shown in figure~\ref{fig:testsetup} and is described as follows:

\begin{itemize}
	\item \textbf{Control Server}: communicates to and controls the initiation and destruction of the virtual environment using \ac{DevOps}.  We use the architecture to test several instances of infected virtual machines.  

	\item \textbf{Hypervisor}: a type 2 hypervisor is used to control resources of the virtual machine.  A type 2 hypervisor is one in which that it is an application that hosts guest operating systems.
	
	\item \textbf{Virtual Machine Store}: a virtual machine store is connected to and pulls the disk image using \ac{DevOps}.  A virtual machine store is one in which several operating system types are available for execution.  
	
	\item \textbf{Virtual Environment}: a virtual environment is comprised of 10 virtual machines that are initiated and destroyed in the Timed Execution period for approximately 1000 data sets.
	
	\item \textbf{Results Database}: a database that contains the collected data.  The collected data (CPU, Memory, Drive) utilization is published to the database at a rate of an entry per 0.5 seconds.
\end{itemize}

We setup the experimentation to support a variety of operating scenarios.  These scenarios are:
\begin{itemize}
	\item [] \textbf{Normal Usage}: Ransomware is not executing and the target operating system is not experiencing any sort of attack
	\item [] \textbf{Timed Ransomware Attack}: Ransomware is executing at a static fixed point either in parallel with normal software or after normal software execution
	\item [] \textbf{Randomized Timed Ransomware Attack}: Ransomware is executing in parallel or after other application usage in random time 
\end{itemize}

\subsubsection*{\textbf{Parallel Execution Events}}
\begin{itemize}
	\item[]\textbf{t$_i$}:  the test period where the system is idle. The system sits post boot up and initialization.  The time is random prior to any actions taking place on the system. 
	
	\item[]\textbf{t$_s$}:  simultaneous activity of events.  There are two installation events Mozilla Firefox and the GNU GCC compiler as well as one TAR GNU Zip compression event with an optional ransomware event.
	
	\item[]\textbf{Time Buffer (t$_b$)}: a time buffer which allows for completion of activities prior to machine destruction. 
\end{itemize} 

\subsubsection*{\textbf{Sequential Execution Events}}
\begin{itemize}
	\item[]\textbf{t$_i$}:  the test period where the system is idle. The system sits post boot up and initialization.  The time is random prior to any actions taking place on the system. 
	
	\item[]\textbf{t$_a$}:  application execution time period event.

	\item[]\textbf{t$_r$}:  ransomware execution time period event.
	
	\item[]\textbf{Time Buffer (t$_b$)}: a time buffer which allows for completion of activities prior to machine destruction. 
\end{itemize}

\subsection*{H$_0$ Null Experiment | Normal Usage of the system}
\begin{figure}[h]
	\centering
	\begin{subfigure}{
			\includegraphics[width=0.47\linewidth]{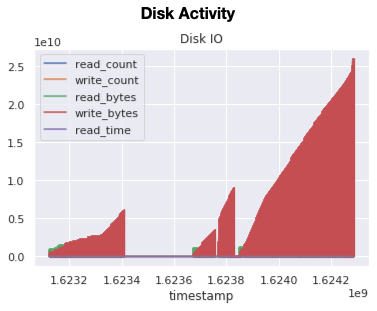}}
	\end{subfigure} 
	\begin{subfigure}{
			\includegraphics[width=0.47\linewidth]{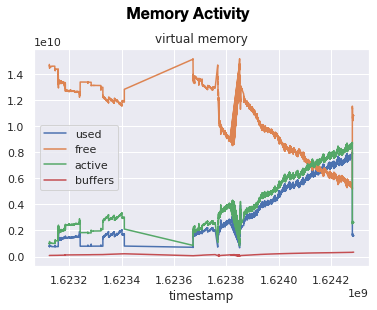}}
	\end{subfigure}    
	\hfill
	\caption{Normal memory and disk utilization as a user is using the system via performing activities such as web browsing and using OpenOffice in a Linux distribution.}
	\label{fig:normaltimeseries}
\end{figure}

In this test set we perform a series of actions from using the system to installing applications while collecting resource data. There are three distinct time periods where the events are illustrated in figure~\ref{fig:normaltimeseries} and referenced in table~\ref{tbl:experimentsetdecomp} as $H_0$.            

The purpose of this experiment set is to characterize the system without ransomware activity.  Random events were executed in parallel within a random event time $t_s$ where $t_e$ in table~\ref{tbl:experimentsetdecomp} represents events $t_i, t_s, t_b$. 

As expected CPU, Memory and Disk utilization follows a pattern of activity consistent of normal system usage.  We do not notice a spike in activity and notice a series of downtime. As previously mentioned, in this experiment ransomware was not present on the system.

\subsection*{$H_{1a}$ Parallel Ransomware Execution}
\begin{figure}[h]
	\centering
	\begin{subfigure}{
			\includegraphics[width=0.47\linewidth]{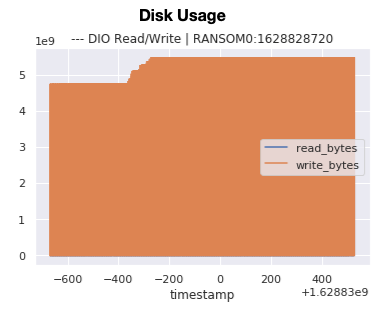}}
	\end{subfigure} 
	\begin{subfigure}{
			\includegraphics[width=0.47\linewidth]{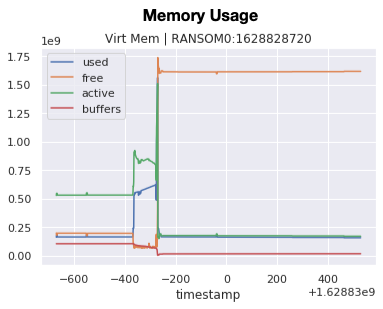}}
	\end{subfigure}    
	\hfill
	\caption{Application usage with parallel execution of ransomware.  In our initial observation we noticed spikes in disk and memory utilization as the processes begin to execute.}
	\label{fig:parallel}
\end{figure}

Shown in figure~\ref{fig:parallel} is the experiment with parallel application execution with Ransomware. The time periods $t_i, t_s, t_b$ start times are chosen randomly.  Table~\ref{tbl:experimentsetdecomp} set expression $H_1a$ describes the experimental time periods and forked application execution.  As in with the null experiment $t_b$ allocates time for the experiment to complete execution.

Applications such as Mozilla Firefox and GCC represented the application installation process while TAR-GZIP represented an application execution process of archiving and compression of files which would heavily utilize CPU, memory and disk activity with added Gonnacry ransomware execution.

%

\subsection*{$H_{1b}$ Sequential Execution With Random Time Periods Ransomeware}

\begin{figure}[h]
	\centering
	\begin{subfigure}{
			\includegraphics[width=0.47\linewidth]{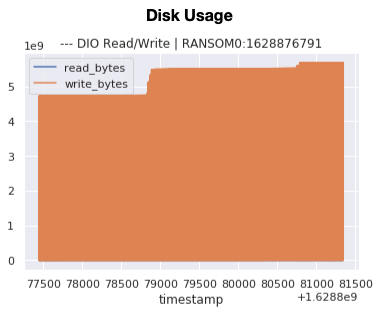}}
	\end{subfigure}
	\begin{subfigure}{
			\includegraphics[width=0.47\linewidth]{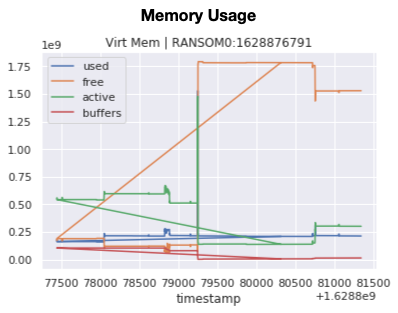}}
	\end{subfigure}    
	\hfill
	\caption{Sequential Execution}
	\label{fig:sequential}
\end{figure}

Shown in figure~\ref{fig:sequential} is the experimental tests with randomization start time and random start times for sequential application and compression execution as well as random start time for ransomware execution.  The time periods are $t_i, t_a, t_r, t_b$ the time buffer $t_b$ is statically set to 5 minutes to allow for process termination.

As with the other experiments as shown in table~\ref{tbl:experimentsetdecomp}, 40 experiments were conducted with 4 instances per experiment running.  The time periods were random.  Execution order had the idle time period randomly generated, then application then executed for a random time period and finally ransomware executed in a random time period.

\section*{Analysis}

\subsection*{Experiment Data set}
The collected experimental data was comprised of CPU utilization, computer memory usage, network activity, and disk utilization. Each dataset was time tagged and provided a unique identifier to identify the specific machine. 

\begin{figure}[h]
	\begin{center}
		\includegraphics[width=1\linewidth]{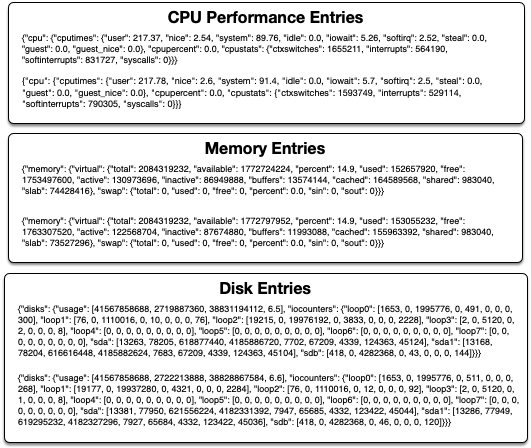}
	\end{center}
	\caption{Sample captured data that comprises only two entries in the respective database tables}
	\label{fig:sampledata}
\end{figure}

In figure~\ref{fig:sampledata} illustrates the type of data that was collected and stored in the database.  The data is stored as a key value pair set of time tagged data. 

The virtual machines that were experimented on could be differentiated by their ID and unique creation time stamp.  As briefly described and shown in figure~\ref{fig:sampledata} the details provides the fidelity needed to characterize a virtual machine instance.

The data set is comprised of $134$ features, Mathematically the data maybe viewed as a temporal sequence of states, i.e., $X[k] \in {\Bbb R}^{134}$.  
This data constitutes the main set of features collected and having various types such as raw counts, percentage utilization, utilization indexes.  The sample rate varies between one and two samples per second. 

\subsection*{Learning Ransomware Behavior} 
To identify executing ransomware, using the data monitor as designed, we create models focused on observing change points to system loads consistent with the requirements of known ransomware samples.  The main difficulty of this problem is that the monitor sees only system load data, rather than more narrowly focused process data.  The challenge is to tease out a signal of an executing ransomware process among all of the other processes that maybe running simultaneously.  

We summarize the learning procedure.
Let $X[k]$ be the sample (system state) at time $k$.  
Since the sample includes data of $N$ features we have $X[k] \in {\Bbb R}^N$.  
The computer monitor which records $M$ samples therefore generates data matrix:
$X_{M,N}$.
We first considering a data transform that characterizes the system utilization changes expected if a ransomware process begins execution.  This data transform involves normalizing the rates of monitored data, and then forming ratios of rates (as a type of comparative homogeneous coordinates):
$$ 
X_{M,N} \rightarrow C_{M-1,N}.
$$ 
Notice the loss of one row due to the first difference procedure. 

Next we define a supervised learning model implemented with ada-boost stumps,which is a type of decision tree classifier.  To specify a sampling method for model training, we define a {\it decision gate}.  The {\it decision gate} governs how time tagged samples from an execution trace will be incorporated into model estimation.   We consider the standard cross fold validation on training and testing sets, but we also create and describe alternative tests aimed to evaluate the uses of decision gates to identify malware.
To test and evaluate the utility of the decision gate for online ransomware detection we design an experiment with populations of test-positive and test-negative traces, and evaluate if the decision gate can reliably and accurately determine the difference given only limited training.  
We summarize the accuracy and how rapidly the classifier can detect ransomware.  Finally we conclude with insights and surprises which this research yields.


\subsection*{Change Point Detection}
Within our model, running processes use resources at various rates, for example at time point $k$, {\it sda1.write.count} increases at a rate proportional to:

$$ r^a = \sum_i^n r_i^{a}(k) f_i(k), $$

with $i$ ranging over the various processes, and $r_i^a(k)$ the rate of process $i$'s utilization of resource $a$; e.g., {\it sda1.writes} at time point $k$, and also $f_i(k)$ the fraction of unit time for which process $i$ is scheduled to run (time proportion to utilize resources).

Assuming equal time sharing among processes (i.e.,  $f_i(k) = 1/n(k)$, with $n(k)$ the number of processes running at time $k$), and letting $A =  \frac{1}{n} \sum_i^n r_i^{a}(k) $, the addition of a new process (process $n+1$) adjusts the rate (utilization of resource $a$) as follows: 
$$
 A  \rightarrow  \frac{n}{n+1} A + \frac{ r_{n+1}^a }{n+1}  
$$

Similarly for resource index $b$, 
$$B \rightarrow \frac{n}{n+1} B + \frac{ r_{n+1}^b }{n + 1}$$ 
and the ratio of index change is:
\[
\ \ \ \ \ \ \ \ \ \ \ \ \ \ \ \ \  \frac{A}{B} \rightarrow \frac{ nA + r^a }{n B + r^b }.\tag{J} \label{eq:J}
\]
This last transformation of relative rate change provides a change point detection target for search.  Further note, that by observing the change of rates resulting from executing ransomware in isolation, (i.e., that is $0 \rightarrow r^e$), we can extended the observation to a set of expected jumps for any system conditions consistent with a specific process starting execution.   Importantly, the training performed within a single system environment can potentially be extended to other system environments thus enabling {\it single shot training}.

These transitional jumps will be incorporated into test/train data transformation, and then provided as features for learning by the ada-boost-stumps classifier.   Our data transformation method outline in algorithm \ref{alg:cap} has three main steps.  The first step is an affine transformation mapping the data matrix into the positive orathont bounded by one as $A$.  Next to reduce noise due to a limited sampling rate, step two applies exponential smoothing.  The reintegrated signal $B$ is smoother, which should facilitate an more persistent decision boundary if applied over subsequent states of the trace.  Data from $B$ can be considered monotonic indices of resource use, as there values increasing in proportion to system utilization.  The third step forms ratios of the monotonically increasing indices as $C$, this step allows us to compare rates of use $r^a, r^b$ and cancels out the distortions of normalizing data in step 1.  We use only $n$ ratios as they can form a multiplicative bases $\frac{r^a}{r^c} = \frac{r^a}{r^b} \frac{r^b}{r^c}$.  

\RestyleAlgo{ruled}
\SetKwComment{Comment}{/* }{ */}

\begin{algorithm}[hbt!]
\caption{Ratio of rate Transform}\label{alg:cap}
\KwData{$X, \delta$, $X$ is $ m \times n$ rows are time samples, $\delta \in [0,1]$ discount factor.  }
\KwResult{$C$ homogeneous data ratio of rates}
$\Delta X = X[2:n,:] - X[1:(n-1),:]$\;
$M,m = ( \max{( \Delta X )}, \min{(\Delta X)} )$\; 
$A = \frac{1}{M-m} \left( \Delta X + m * \text{ones}( n-1, m ) \right)$\;
$B = A$\Comment*[r]{1: A in unit orathont }
$k \gets 2$\; 

\While{$k \leq n-1$}{
  $B[k,:] = \delta * B[k-1,:] + (1-\delta) A[k,:]$\;
  $k \gets k + 1$\;
}
$C = B$\Comment*[r]{2: B exp smooth}
$j \gets 1$ \;
\While{$j \leq m$}{
$C[:,j] = B[:,j] ./ B[:(j+1)\% m]$\;
$j \gets j + 1$\;
}
$C$ \Comment*[r]{C ratio of rates data}
\end{algorithm}

The transformation is aimed to tease out a measurable change in rates, under the assumption that all other processes are in rate equilibrium, if a new process starts between sample $k$ and $k+1$, the effect leads only to increased rates of use in the rows (time samples) of $B$ before (and including step) $k$ when compared to time samples immediately after $k$.  These rate adjustments cause the ratios of rates time samples (rows of $C$) to undergo a jump process moving to one ratio equilibrium to another.  

While we can control for the environment (set of other running process) during training, accounting for all potential environments where the classifier is expected to operate is an enormous problem.  By characterizing the step by step changes to any environment that are consistent with ransomware we hope to address the sizable problem of accounting for all possible environments where the detector must operate.  While we make the inaccurate assumption that environments are in equilibrium when ransomware starts, we note first that sampling rate is critical (as a Nyquist Sampling barrier argument would support) and second that a smoothing factor $\delta$ in algorithm \ref{alg:cap} is helpful toward isolating significant shifts. 

\subsection*{The Decision Gate Model} 
With the data transform above, we proceed by describing the supervised learning framework.  Here we define both the hypothesis space and the learning objective.
Our data consists of {\it traces}, with each trace being a sequence of states measuring relative rates of utilization across the various monitored system aspects.  
We define an {\it Epoch} to be any sub-sequence of a trace defined by a start and end, i.e., a contiguous set of samples.  
In our training set we will have tags or certain knowledge of when malware was started, thus we can tag the epoch of infection.  For each trace $t$ we record the start time for the ransomware as $s(t)$, we tag the time samples of trace with labels:
$$
\Phi_t(k) = \begin{cases} 0 \text{ if } k < s(t) \text{ benign Epoch } \\ 1 \text{ otherwise for the infected Epoch } \end{cases}
$$

Next we consider a sampling strategy for traces in order to learn features characterizing an epoch.  We continue by defining the {\it gate} function:

$$
G( k ; \alpha, \beta ) = \begin{cases} 1 \text{ \ if } k :  \alpha \leq k \leq  \alpha + \beta \\ 
0 \text{ \ otherwise } \end{cases}  
$$

For a trace $t$ that executes malware (starting at time $s(t)$, our learning method will sample states form the trace by using the gate function $G( k , s(t) + \alpha, \beta )$.  This sampler allowing $\alpha$ steps to occur after the malware starts and then sampling an epoch of size $\beta$.  The reasoning is that we wish to capture our rate ratios in steady state, while $\beta$ large is desirable, being too large runs the risk that transitions between steady state may confounding the classifier.  
Alternatively to sample the benign epoch we select $b$ samples uniform randomly over $\Phi^{-1}_t ( 0 )$ i.e., the benign epoch. 

Since smoothing induced by the parameter $\delta$ attenuates noise both in the benign epoch as well as in $t$'s Gate $G(k, s(t) + \alpha , \beta )$, we consider it to also be an important parameter for sampling, as increased smoothing could facilitate a larger epoch $\beta$ for observation but may also require increasing an offset $\alpha$ to defray the effects of a rapid change or transition from equilibria. 

{\bf Decision Gate:} We define the {\it decision gate} as the sampling and smoothing parameters $\langle \alpha , \beta, \delta \rangle$.  

{\bf Supervised Learning:} Supervised learning often follows a training/testing framework to including cross-fold validation.  We follow the usual pattern for Ada boost stump classifiers trained on decision gates for ground truth traces.  Since the purpose of our technique is intended to be online or real-time ransomware identification we perform a second validation experiment designed to evaluate the merits and suitability for such real-time change point detection capable of attributing a system change to ransomware attack. 

With the decision gate specified, we use a small set of tagged trace data (where ransomware was executed) to train a binary classifier over the feature space.  Using the tagged data we continue by assessing the mean accuracy in 10x cross-fold validation.

Our second experiment designed to asses the use of the classifier for real-time change point detection.   Once formed the binary classifier is tested on a two sets of 36 traces held out from training, but also where the full answer is known.  first the rate of identification for a test positive set (where malware is started and should be detected) is measured.  Next the rate of identification is measured on a test negative data set (where no malware is run, other applications are executed and no detection is desired).  These two test validate the classifier's ability to distinguish infection from benign and could be considered additional testing in a training and testing process.  But we also consider the identifications and their accuracy:   for the malware identified (in the test positive case) we measure how accurately the detection is by capturing the time between malware starting $s(t)$ and the detection.  Said differently this additional measure assesses the speed at which the classifier can identify the malware once started.

{\bf Supervised Learning and Cross Fold Validation:}

The learning objective is to distinguish the $\delta$ smoothed states expressed in the decision gate from a comparable sized set of states randomly distributed over the benign epoch. 
We train a adaboost decision stumps classifier implemented in Julia Decision Trees.  
For traces which simultaneously execute applications, we additionally sample from a gate structure (with same parameters) for the application startup.  
This additional sampling is tagged as benign and ensures that our classifier is focused on malware activities as opposed to just any activity starting within an idle background.



To assess accuracy (as a form of statistical power) of the stumps decision classifier we use cross fold validation which essentially measures mean accuracy across multiple train/test experiments.  Specifically in a $k-$fold validation the tagged data is randomized into $k$ equal sized portions to form $k$ experiments.  In the $j^{th}$ experiment, all but the $j^{th}$ portion will be combined to form the training set and the resulting classifier is tested against on the hold out (portion $j$).  
The statistical power measures are collected over the $k$ experiments (folds) and summarized statistically to assess accuracy, the experiments also offer a means to observe stability of the resulting classifiers, transitional parameters (of the decision gate) and other considerations such as model complexity (depth and interpret-ability of the decision structure). 

\begin{figure}[!h]
     \centering
     \begin{subfigure}{
         \includegraphics[width=\linewidth]{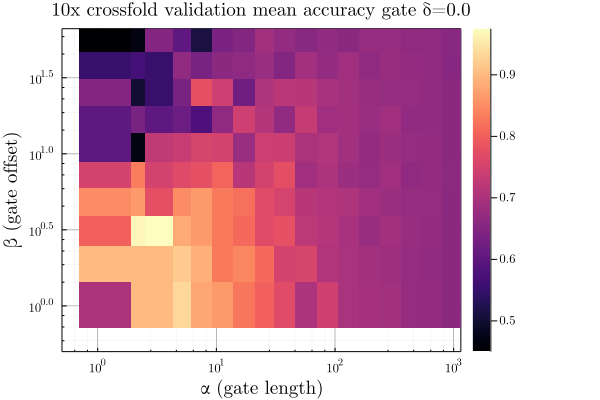}}
     \end{subfigure}
     \\
     \begin{subfigure}{
         \includegraphics[width=0.47\linewidth]{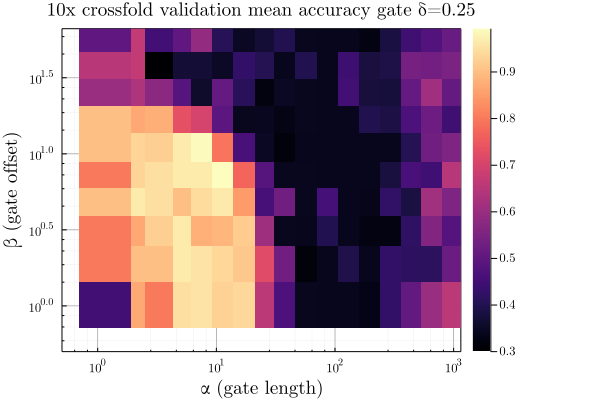}}
     \end{subfigure} 
     \begin{subfigure}{
         \includegraphics[width=0.47\linewidth]{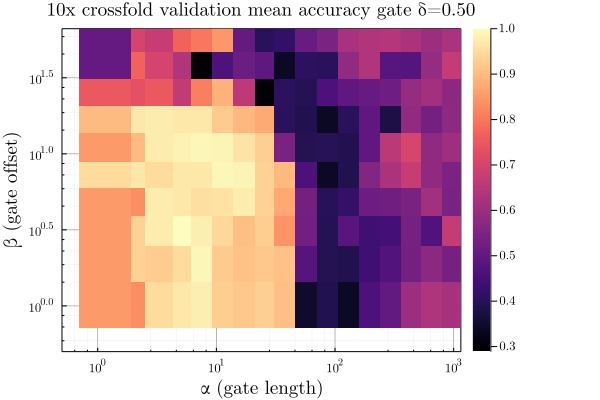}}
     \end{subfigure}
     \\
     \begin{subfigure}{
         \includegraphics[width=0.47\linewidth]{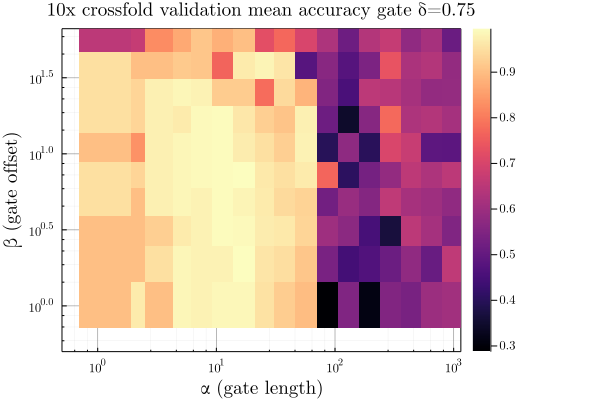}}
     \end{subfigure}
    \begin{subfigure}{
         \includegraphics[width=0.47\linewidth]{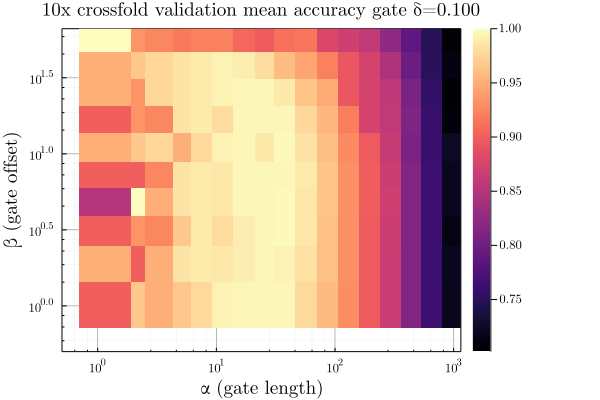}}
     \end{subfigure}      
     \hfill
        \caption{Accuracy for the change point detection as a function of the decision gate parameters.  In each diagram the gate length $\alpha$ is plotted on the x-axis (logarithmic scale), the gate offset $\beta$ on the y-axis (logarithmic scale), and intensity values illustrate the mean accuracy of the adaboost stumps classifier constructed from the decision gate.  The same plot is illustrated for values of $\delta$ ($\delta = 0.25$, $\delta = 0.5$, $\delta = 0.75$, and $\delta = 1.0$ in raster order for the lower four heatmaps).  Note that the effect of smoothing $\delta$:  a slight trade off in accuracy for a larger sweet-spot.  Interestingly the case of $\delta=0$ appears to differ from the non-zero cases which feature a noticeable decrease in accuracy at a given length.    }
        \label{fig:crossfold}
\end{figure}

\subsection*{Learning procedures tested in real-time detection:}
Pursuant to our original goal, we evaluate the change point detection technique on a corpus of additional test data.  Letting $\langle \alpha, \beta, \delta \rangle$ describe a gate structure we construct decision trees from the specified gate on four instances of training data.  
Next we use the constructed decision gate and apply to un-observed traces from two sets:  a test-positive set of $40$ traces (including our four training instances) and a test-negative data set of $36$ traces.  The test-positive traces run ransomware and we have ground truth time-stamps for when,  the test-negative traces do not run ransomware.  A variety of applications are executed in both test-positive and test-negative data sets including installers and compression.  All details are whit-held from the learner,  aside from the four traces used for training, the learner will have to rely on the decision surface constructed to make assignments in the 76 traces presented to it.  
The adaboost decision stumps classifier will be used to measure the probability of ransomware for each state.  If the probability of ransomware exceeds the threshold $\tau$ the classifier will flag a real-time identification.   
To return to the goal, and for correctly identified ransomware, we characterize the delay in identification with a statistical histogram measuring number of samples before an $\tau$ assignment can be made.

\begin{figure}[h!]
     \centering
     \begin{subfigure}[a: Rates of classification for $\tau$ ]{
         \includegraphics[width=\linewidth]{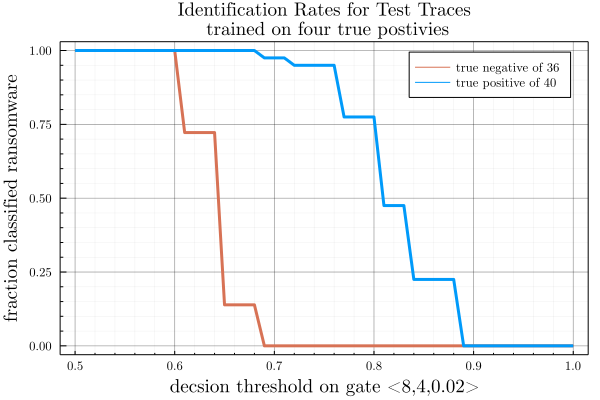}}
     \end{subfigure}
     \hfill
     \begin{subfigure}[b: Detection delays $\tau=0.75$]{
         \includegraphics[width=\linewidth]{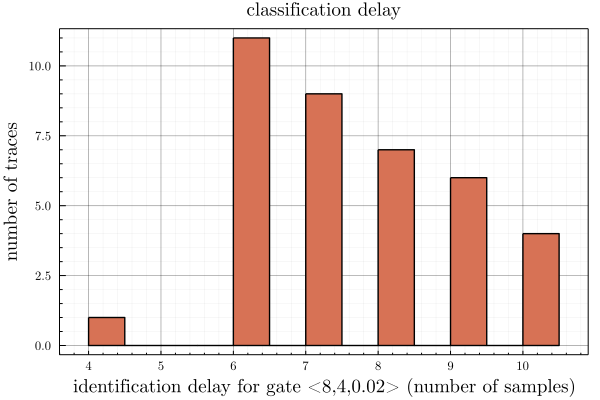}}
     \end{subfigure} 
        \caption{Training a classifier - Inferring Ransomware Attacks. }
        \label{fig:gate1}
\end{figure}

In figure \ref{fig:gate1}, we illustrate the results of using gate $\langle 8, 4, 0.02 \rangle$, in two graphs.  In the top graph we show the rates of ransomware identification for the test-positive set (blue) and test-negative set (orange) over $\tau$ the decision parameter.  
Note that above $0.7$ the classifier correctly filters out non-malware traces and captures only ransomware, up to $\tau 0.884$ where the decision criteria becomes to stringent.  
One can observe the region for $\tau$ where the classifier splits cases.  While there is no perfect value for $\tau$ (in this case), the overall operating characteristics appear promising.
In the histogram we illustrate the delay for ransomware classification.  Choosing $\tau = 0.75$ we correctly identify 38 of 40 traces and make no false positive errors.  One trace is identified in 4 samples, all 38 are identified within 10 samples, which is less than 10 seconds. The histogram suggests a mode of $6$ samples or roughly 4 seconds.

\begin{figure}[h!]
     \centering
     \begin{subfigure}[c: Rates of classification for $\tau$]{
         \includegraphics[width=\linewidth]{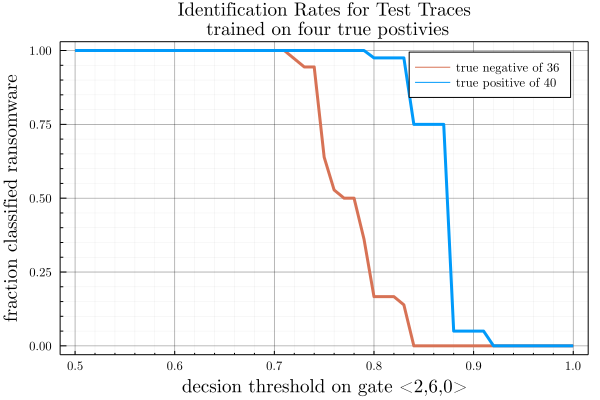}}
     \end{subfigure}
     \hfill
     \begin{subfigure}[d: Detection delays $\tau=0.84$]{
         \includegraphics[width=\linewidth]{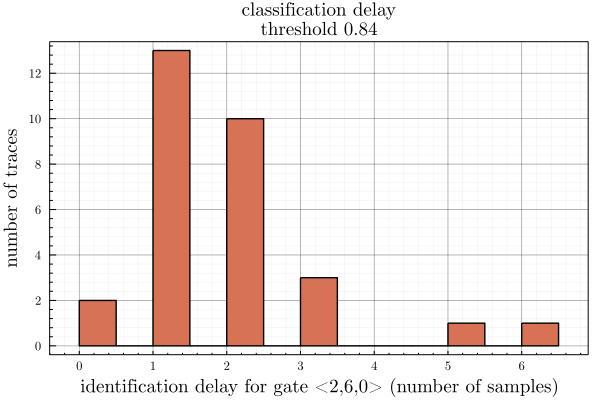}}
     \end{subfigure} 
        \caption{Relationship of gate parameters to classifier characteristics.}
        \label{fig:gate2}
\end{figure}

In figure \ref{fig:gate2}, we illustrate the results of using gate $\langle 2, 6, 0.0 \rangle $.  Similarly we observe a region for decision parameter $\tau$ that effectively identifies the ransomware change characteristics.  Although the area between the curves is less, suggesting less robustness than our first example, there still exists a clear opportunity for identification.  
In the histogram plotted for $\tau=0.84$ one can see that the classifier acts quicker than the classifier of the prior discussion.  This is due to there being no smoothing, so one state observation can suffice if the features observed are well within the decision surface.   This classifier has a mode of one sample ( about 0.6s ).  One sample was identified with a single state observation.  All 30 (of 40) ransomware traces  were recognized within 6 samples (about 4 seconds).

\begin{table*}[!ht]
	\caption{Learning procedures considered}
	\label{tbl:ML}
	\centering
	\begin{tabular}{|l|l|l|}
		\hline
		\multicolumn{1}{|c|}{\textbf{Method}} & \multicolumn{1}{c|}{\textbf{Description}} & \multicolumn{1}{c|}{\textbf{Summary}}                                                                                                                            \\ \hline
		\textbf{ PCA }    & 
		Using decomposition techniques  upon the data matrix
		& Detection of ransomware in real-time \\ & & by subspace projection      
		\\ \hline
		\textbf{Decision Tree}  & Use of binary decisions to qualify infection & 
		Low complex decision surface \\
		\hline
	    \textbf{MDPs}  & Use of state to state transitional structures & Requires state estimation      \\ \hline
	\end{tabular}
\end{table*}

{\bf Evaluation of merits for ML classifiers:}

The adaboost stumps decision classifier is capable of detecting the start of ransomware within the system monitor described.  We found that the adaboost stumps method seemed more reliable and stable than the standard decision tree. 
We found the search for effective gate parameters to be difficult with hit-or-miss sensitivities and repeatably issues (for regular decision trees)
These problems were abated by use of the adaboost stumps classifier in replacement of regular decision trees and an exhaustive parameters sweeping search as seen in cross-fold validation results figure (\ref{fig:crossfold}).  We suggest the use of parameter sweeping search for gate parameters to guide the search for effective parameters and also to verify the structuralism in the cross fold validation mean accuracy measure  
The results of our real-time ransomware identification experiment suggest that the method can potentially be tuned to form effective ransomware identifiers.    Still the model of decision gates is modular, and can easily be extended to its own ensemble model of either nested or serial decision gates, a problem we intend to explore in future work.

Referencing Figure~\ref{fig:gate1}: How the trained classifier infers ransomware attacks.  We train an Ada-boost decision stumps model on four test-positive traces sampling according to the gate parameter.  Next we then examine the classifier's ability to distinguish two populations of test traces: test positive (executes applications and ransomware) and test negatives (executes applications but not ransomware).  Having only experience on training data, the classifier will not know if a trace is positive or negative for ransomware.  It makes inferences by calculating a probability that the trace is executing malware at each time-point, detection occurs when the model probability exceed $\tau$.  In [a] we measure the rates of classification for the two populations.  An ideal classifier separates the two populations for some values of $\tau$, for example detecting all true positives and never detecting any true negatives. When detection occurs, we can evaluate how accurately the result is by measuring the number of samples between the ransomware start time (also withheld from the classifier) and the classifier's first detection time.  These detection delays are plotted as a histogram in [b] for $\tau = 0.75$.

Referencing Figure~\ref{fig:gate2} and continuing from figure \ref{fig:gate1}, we illustrate the relation of gate parameters to classifier characteristics, noting sensitivities and trade-offs.  In \ref{fig:gate1} [a,b] the gate parameters prescribe a capture of $4$ samples after a delay of $8$ samples (after the ransomware is executed).  The training data uses only four states ($8$ after the ransomware starts) to train the model, it uses a similar sample size for application data and idle time.  Further exponential smoothing is applied with $\delta=0.2$.  In [c,d] the gate is a set of $6$ states only $2$ samples after the execution event - with no exponential smoothing.  Notice the decrease of area between the two curves in [c] vs [a] suggests that waiting and smoothing may enhance robustness, but also slightly increases detection delay.
\section*{Related Works}
Our experimental setup can be considered a type of Malware Run-time Analysis System, also sometimes referred to as a Sandbox Run-time System, Self Monitor, or Behavior Test Harness.  There have been many published projects which follow the same approach to combine run-time systems that collect data to analyze signatures or behavior allowing the identification of Malware.  Three early works include using clustering, sub-graph matching, and machine learning to detect malware include \cite{bayer2009scalable, park2010fast, rieck2011automatic, ransommitigation}. 
The prominent architectural features of our approach is consistent with the related work and consists of:  first, a run-time component that runs malware in a contained way in order to observe and record event sequences; and second, a component to ingest the data generated from the run-time system to generalize rules to detect known malware, thereby enabling a type of inductive reasoning.  
While there is some interest in the difficulty of classifying malware (and methods to translate the problem to human recognition tasks \cite{vanhoudnos2017malware}), one benefit of using the  statistical classifiers or machine learning approach within the second component owes to its fastidious ability to track and identify relevant features from the large number of features observed from the first component.  
As such, and not knowing (a priori) how a malware may reveal its behavioral when run, the use of machine learning is consistent with bertillonage (\cite{davies2011software}) or the collection of a broad range of features as any one feature may prove pivotal to distinguishing a malware behavior from benign background.  
Since these systems are re-engineered for each system, some interest has also arisen on unified model for malware detection, malicious behavior \cite{casey2014agent, bilar2006statistical, ghosh2017malware}.

\section*{Conclusion}
We found that once trained the Machine Learning algorithms are capable of determining the execution of ransomware in a timely manner.  We were able to prove our hypothesis and disprove the null hypothesis through experimentation.  The algorithms are able to detect and potentially have additional capability to mitigate the effects of ransomware.  We also found that additional automated testing is needed in two distinct areas.  First, additional ransomware on operating systems other than LINUX i.e. Microsoft Windows and Apple Macintosh.  Second, improvements in testing with randomization of time periods caused the machine learning algorithms to change resulting in better predictions.  We believe as ransomware continues to evolve so should the testing and detection methods used should also evolve.  

\section*{Acknowledgments}
The authors would like to thank Daniel Koller of \ac{ONR} for sponsoring this work. The authors would also like to the thank the \ac{USNA} \ac{CECSR} group for thoughful discussions in the area of Ransomware research.

\bibliographystyle{IEEEtran}
\bibliography{bibfile}

\end{document}